# Structure Sensitivity in Oxide Catalysis:
# First-Principles Kinetic Monte Carlo Simulations for CO Oxidation at RuO$_2$(111)

Tongyu Wang[1] and Karsten Reuter[1,2,*]

[1]Chair for Theoretical Chemistry and Catalysis Research Center,
Technische Universität München, Lichtenbergstr. 4, D-85747 Garching, Germany
[2]SUNCAT Center for Interface Science and Catalysis, SLAC National Accelerator Laboratory
& Stanford University, 443 Via Ortega, Stanford, CA 94035-4300, U.S.A.

*E-mail: karsten.reuter@ch.tum.de

We present a density-functional theory based kinetic Monte Carlo study of CO oxidation at the (111) facet of RuO$_2$. We compare the detailed insight into elementary processes, steady-state surface coverages and catalytic activity to equivalent published simulation data for the frequently studied RuO$_2$(110) facet. Qualitative differences are identified in virtually every aspect ranging from binding energetics over lateral interactions to the interplay of elementary processes at the different active sites. Nevertheless, particularly at technologically relevant elevated temperatures, near-ambient pressures and near-stoichiometric feeds both facets exhibit almost identical catalytic activity. These findings challenge the traditional definition of structure sensitivity based on macroscopically observable turnover frequencies and allow to scrutinize the applicability of structure sensitivity classifications developed for metals to oxide catalysis.

## 1 INTRODUCTION

Detailed kinetic studies comparing the catalytic activity of different single-crystal facets provide important insights on several accounts. They allow for a straightforward assessment of the structure sensitivity of the catalytic reaction[1-3], which constitutes an important first milestone when aiming to relate the detailed knowledge from model catalysts to the performance of real supported catalysts. They also contribute to a more systematic bridging of the materials gap when employed to analyze data obtained for polycrystalline powders. This holds in particular if microkinetic models established for different facets are suitably combined to address the catalytic activity as a function of size and shape of the active nanoparticles. Recent years have seen the advent of predictive-quality first-principles microkinetic models.[4] Due to the still notable computational costs in obtaining the underlying first-principles data, such work has hitherto largely focused on the study of individual single-crystal facets. As a first step towards a full first-principles microkinetic model of a nanoparticle we here present a detailed first-principles kinetic Monte Carlo (1p-kMC) study of CO oxidation at RuO$_2$(111).

Over the years CO oxidation at RuO$_2$ has developed into a most extensively studied system, originally motivated to rationalize the qualitative activity differences of Ru catalysts in ultra-high vacuum and ambient conditions.[5-7] Almost all of the single-crystal work has thereby been focused on RuO$_2$(110), which forms upon oxidation of the close-packed Ru(0001) surface and which constitutes the lateral facets of RuO$_2$ crystals. As this surface was shown to microfacet into an inactive c(2x2)-RuO$_2$(100) phase under oxidizing conditions[8], recent experimental[9] and theoretical[10] work has pointed at a possibly prominent role of apical RuO$_2$(111) facets for the long-term catalytic activity in such feeds. In the present work we therefore focus on this facet and compare extensively to the established 1p-kMC model for CO oxidation at the hitherto primarily investigated RuO$_2$(110) facet[11-14]. In order to allow a most meaningful comparison, we thereby employ exactly the same microkinetic modeling approach, namely 1p-kMC[15], and the same density-functional theory (DFT) exchange-correlation functional, namely the generalized-gradient functional due to Perdew, Burke and Ernzerhof (PBE)[16].

Previous work on structure sensitivity at metal catalysts has emphasized the role of electronic effects due to a different degree of under-coordination of surface atoms, as well as the role of geometric effects due to different bonding configurations.[17,18] At the structurally more complex oxide surfaces we find that the spatial distribution of the active sites and concomitant diffusion limitations constitute a third important factor. We identify qualitative differences in all three respects in the CO oxidation at $RuO_2(111)$ and $RuO_2(110)$. As such one would clearly classify the reaction as structure sensitive. However, the catalytic activities of the two facets peak with very similar maximum turnover frequencies (TOFs) at different reactant partial pressure ratios. This highlights that care has to be taken when assessing a potential structure sensitivity merely on the basis of comparable catalytic activity in a restricted range of feed conditions: At least for $RuO_2(111)$ and $RuO_2(110)$ near-ambient feed conditions can be found where depending on the exact partial pressure ratio both facets exhibit either virtually identical or largely differing TOFs.

## 2 METHODOLOGY

We evaluate the kinetics of the reaction network using 1p-kMC simulations.[4,15] For steady-state reaction conditions defined by temperature and reactant partial pressures, ($T, p_{O_2}, p_{CO}$), the central outcome of such simulations are the overall catalytic activity (measured as TOF in product molecules per area and time) and average coverages at the surface. In contrast to prevalent mean-field rate equation based microkinetic simulations, 1p-kMC thereby fully accounts for the correlations, fluctuations, and explicit spatial distributions of the reaction intermediates at the catalyst surface.[13] This allows to analyze in detail the occurrence and contribution of any elementary process or local surface configuration within the entire reaction network. The input required for 1p-kMC simulations are a list of all elementary processes in the reaction network and their respective rate constants. We evaluate the latter using DFT and transition state theory (TST).[12] Additionally required is a lattice model that specifies the geometric arrangement of the individual surface sites involved in the reaction network. In the following we first summarize this lattice model and the list of elementary reactions considered, and then describe the computational procedure to obtain the first-principles rate constants.

*2.1 Lattice model and elementary processes*
The lattice arrangement of rutile $RuO_2$ along the [111] direction can be seen as a stacking of ($RuO_4$)-Ru bilayers as illustrated in Fig. 1a. Each bilayer is laterally displaced from the one underneath until the sequence repeats itself after four bilayers. Each $RuO_4$ plane exhibits three non-equivalent O atoms, which together with the Ru plane leads to a total of four possible $RuO_2(111)$-(1x1) terminations.[10] We choose the most stable O-poor termination involving the $RuO_4$ plane as basis for the 1p-kMC lattice-model, as this then naturally accommodates all other O-terminations as a consequence of O adsorption processes. As shown in Fig. 1b the chosen O-poor termination exposes three under-coordinated Ru atoms per surface unit-cell: One threefold coordinated Ru atom (labeled as $Ru_2$) and two fivefold coordinated Ru atoms (labeled as $Ru_1$ and $Ru_3$, respectively). Systematically exploring O and CO adsorption at all high-symmetry sites, our DFT calculations identified three possible adsorption sites close to these Ru atoms: A bridge site between $Ru_1$ and $Ru_2$ (labeled as site $Ru_1Ru_2$), a site atop of $Ru_2$ (labeled as $Ru_2$) and a bridge site between $Ru_2$ and $Ru_3$ (labeled as site $Ru_2Ru_3$). These are exactly the sites that would be occupied by O atoms in the continuation of the bulk stacking sequence, i.e. we did not find any additional adsorption sites stabilized as a consequence of the lattice truncation at the surface.

Figure 1c depicts the lateral arrangement of these three adsorption sites. Their linear arrangement in form of a $Ru_1Ru_2 - Ru_2 - Ru_2Ru_3$ site chain enables two-site processes like dissociative $O_2$ adsorption, associative $O_2$ desorption, O and CO diffusion, as well as Langmuir-

Hinshelwood (LH) type CO oxidation at and between directly neighboring site pairs, but not at or between the most distant $Ru_1Ru_2$ and $Ru_2Ru_3$ site pair. Even more intriguing is the large geometric distance of any of these three sites to any site in neighboring surface unit-cells. As shown in Fig. 1c the closest such distances are between a $Ru_2Ru_3$ site in one cell and a $Ru_1Ru_2$ in the nearest-neighboring surface unit cell (3.30 Å), and between a $Ru_2$ site and a $Ru_1Ru_2$ site in the nearest-neighboring surface unit cell (4.69 Å). We tested for two-site processes involving these site pairs, but always obtained prohibitively large barriers. Only diffusion processes between $Ru_2$ and $Ru_1Ru_2$ sites across unit cells led to reasonable barriers and are accordingly considered in the 1p-kMC model.

Essentially, this thus leads to a lattice model for CO oxidation at $RuO_2(111)$ that has more of a molecular character than that of an extended surface network. Within the ensemble of three sites $Ru_1Ru_2$–$Ru_2$–$Ru_2Ru_3$ within one surface unit-cell we consider non-concerted adsorption, desorption, diffusion and reaction processes, with two-site processes restricted to nearest-neighbor pairs within the three-site chain. Specifically, dissociative $O_2$ adsorption can occur on empty $Ru_1Ru_2$–$Ru_2$ or $Ru_2$–$Ru_2Ru_3$ pairs, molecular CO adsorption on any empty site, and O/CO diffusion to a nearest-neighboring empty site. Desorption processes are modeled as time-reversed counterparts of the adsorption processes. LH oxidation reactions are possible between O and CO occupying nearest-neighbor sites and lead to an immediately desorbing $CO_2$ product molecule. Eley-Rideal (ER) oxidation reactions in form of gas-phase CO scattering are possible with O adsorbed at any of the three sites and lead equally to an instantaneously desorbing $CO_2$. The only additional processes that can connect sites in adjacent surface unit-cells are diffusion processes between $Ru_2$ and $Ru_1Ru_2$ sites. At the feed conditions discussed below, these diffusion processes have only a quantitative effect though. Switching them off in the 1p-kMC simulations left all conclusions put forward below intact and led to TOF changes at peak activities of the order of five. In practice, the lattice model can therefore be seen as a finite three-site model.

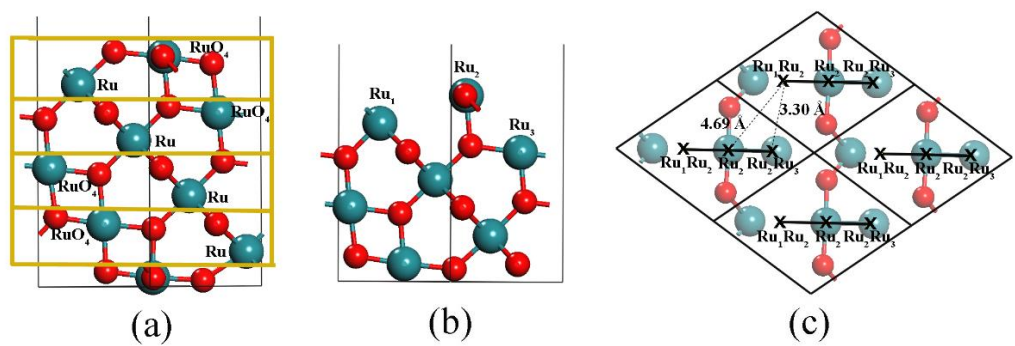

Figure 1. (a) Side view illustrating the stacking sequence of rutile $RuO_2(111)$ with every ($RuO_4$)-Ru bilayer marked with a yellow box to highlight the repeat sequence after four bilayers. (b) Side view of the most stable O-poor termination chosen as basis for the 1p-kMC lattice model (see text). (c) Top view showing four $RuO_2(111)$ surface unit-cells. Additionally marked are the three surface sites ($Ru_1Ru_2$, $Ru_2$, $Ru_2Ru_3$) considered in the 1p-kMC model and the shortest geometric distances across surface unit-cells. Ru atoms are shown as large blue spheres, O atoms as small red spheres.

*II.2 First-principles rate constants and computational details*
The calculation of the first-principles rate constants follows the approach put forward by Reuter and Scheffler[12]. This approach relies on kinetic gas theory to determine the rate constants for adsorption processes and ER reaction processes. Rate constants for time-reversed desorption processes are determined through detailed balance. For bound-to-bound transitions like surface diffusion processes or activated LH reactions harmonic TST is applied. This reduces the first-principles input necessary to calculate the rate constants essentially to binding energies and reaction barriers. These energetic parameters are obtained from DFT using the plane-wave code CASTEP[19]. Electronic exchange and correlation is treated at the level of the PBE generalized gradient functional[16]. The core electrons are described by standard library ultrasoft pseudopotentials, while the valence electrons are expanded in a plane-wave basis set with a cut-off energy of 450 eV. The $RuO_2$(111) surface is modeled with a 7-bilayer slab with the bottom three layers fixed to represent the bulk structure, and a vacuum separation exceeding 10 Å. Reciprocal space integrations are carried out on a (4x4x1) Monkhorst–Pack grid for a (1x1) surface unit-cell. All adsorption geometries are fully relaxed until residual forces are below 50 meV/Å. With this computational setup the binding energies used to determine (thermodynamic) desorption barriers are converged to within 50 meV. We furthermore validated that this setup provides a binding energetics that is fully consistent with the full-potential approach that was employed for the $RuO_2$(110) 1p-kMC model[12]. Explicit surface barrier calculations for LH reaction and diffusion processes were carried out with the climbing-image nudged elastic band (CI-NEB) method[20] as implemented in the ASE environment[21] and using eight images between the known initial and final states. Activation barriers for ER reaction and adsorption processes were obtained through reaction coordinate scans of the potential energy surface. As reaction coordinates we employed constraints on the vertical distance from the surface or in case of ER also the C-O distance between the impinging CO and the adsorbed O atom.

Table 1. $O_2$ and CO desorption barriers (in eV). The different rows specify the sites out of which desorption occurs, while the different columns indicate the occupation of the other site(s) within the surface unit-cell (empty (e), O or CO). The variation over the different columns thus reflects the lateral interactions with nearby adsorbed species.

| $O_2$ desorption barrier | | | | | | | | | |
|---|---|---|---|---|---|---|---|---|---|
| | E | | | O | | | CO | | |
| $O_2$@$Ru_1Ru_2$–$Ru_2$ | 2.96 | | | 3.30 | | | 2.94 | | |
| $O_2$@$Ru_2$–$Ru_2Ru_3$ | 3.22 | | | 2.35 | | | 2.44 | | |
| CO desorption barrier | | | | | | | | | |
| | e e | e O | e CO | O e | O O | O CO | CO e | CO O | CO CO |
| CO@$Ru_1Ru_2$ | 1.54 | 1.42 | 1.71 | 0.47 | 0.76 | 0.01 | 0.90 | 0.82 | 1.04 |
| CO@$Ru_2$ | 1.88 | 1.90 | 1.59 | 1.01 | 1.32 | 0.85 | 1.24 | 1.30 | 0.92 |
| CO@$Ru_2Ru_3$ | 0.98 | 0.99 | 0.69 | 1.49 | 0.96 | 1.33 | 1.15 | 0.53 | 0.83 |

No appreciable activation barriers were identified for the dissociative $O_2$ and molecular CO adsorption processes within the sites of one surface unit-cell. Within the hole model for adsorption underlying the Reuter/Scheffler approach[12], the sticking coefficients needed to determine the adsorption rate constants are then approximately given by the fraction of all impinging molecules that ends up in the corresponding sites/site pairs. For simplicity, we apply an equi-partition and use ½ for dissociative $O_2$ adsorption over the two site pairs $Ru_1Ru_2$–$Ru_2$ and $Ru_2$–$Ru_2Ru_3$, and ⅓ for the uni-molecular CO adsorption over each of the three sites within the unit-cell. Since adsorption is non-activated, all desorption barriers for the time-reversed processes are given by the thermodynamic binding energies. Due to the highly under-coordinated $Ru_2$ surface atom we calculate significant lateral interactions between the adsorbed species, i.e. the bond strength at the individual sites varies largely with the occupation of the other sites within the surface unit-cell. In contrast, there are only negligible lateral interactions between sites across surface unit-cells. Fortunately, the latter thus generates only a small number of different site occupations for which binding energies are required. For dissociative $O_2$ desorption these are three different binding energies of $O_2$ at the $Ru_1Ru_2$–$Ru_2$ pair, depending on whether the $Ru_2Ru_3$ site is empty, or occupied by O or CO. For dissociative $O_2$ desorption out of the $Ru_2$–$Ru_2Ru_3$ pair this is likewise three binding energies, while for the CO desorption out of the three sites a total of 27 binding energies are required (depending on the occupation of the other two sites). This small number of combinatorial possibilities can still be captured by explicitly calculating the binding energy of every configuration. This allows to exactly treat the large lateral interactions implied by the corresponding numbers compiled in Table 1, as compared to the more common approximate treatment in form of short-ranged lattice-gas Hamiltonians[22-25].

Table 2. Diffusion barriers within one $RuO_2(111)$ surface unit-cell (in eV). The different columns indicate the occupation of the third site within the surface unit-cell (empty (e), O or CO). The variation over the different columns thus reflects the lateral interactions with nearby adsorbed species.

| Diffusion | 3$^{rd}$-site | e | O | CO |
|---|---|---|---|---|
| O | $Ru_1Ru_2 \rightarrow Ru_2$ | 1.15 | 0.82 | 1.11 |
| O | $Ru_2 \rightarrow Ru_1Ru_2$ | 0.54 | 0.25 | 0.00 |
| O | $Ru_2 \rightarrow Ru_2Ru_3$ | 0.66 | 0.35 | 0.15 |
| O | $Ru_2Ru_3 \rightarrow Ru_2$ | 0.32 | 1.18 | 0.76 |
| CO | $Ru_1Ru_2 \rightarrow Ru_2$ | 0.71 | 0.20 | 0.43 |
| CO | $Ru_2 \rightarrow Ru_1Ru_2$ | 1.05 | 0.68 | 0.31 |
| CO | $Ru_2 \rightarrow Ru_2Ru_3$ | 1.32 | 0.65 | 0.79 |
| CO | $Ru_2Ru_3 \rightarrow Ru_2$ | 0.42 | 1.13 | 0.70 |

Lateral interactions in diffusion processes between sites within one surface unit-cell are equally resolved by explicitly calculating the initial and transition state for every possible occupation of the third site. This yields the total of 12 forward and 12 backward diffusion process barriers compiled in Table 2. The large barrier variation obtained for the same diffusion process and varying occupation of the third adsorption site reveals equally large lateral interactions as for the desorption processes. This contrasts previous findings for CO oxidation at Pd model catalysts[23-25], where a rough scaling of initial and transition state energies rendered diffusion barriers largely independent of the local environment. As further illustrated below we attribute this difference to the much higher structural flexibility of the largely under-coordinated $Ru_2$ atom at $RuO_2(111)$, which thus adapts more strongly to nearby bonded adsorbates.

For diffusion processes across surface unit-cells, i.e. from site $Ru_2$ in one cell to $Ru_1Ru_2$ in a neighboring cell, test calculations indicate only small variations of the transition state energy with varying occupation of other sites both in the original and in the destination surface unit-cells. Diffusion barriers thus essentially vary only with changes of the initial state energy, i.e. with the binding energy of the diffusing species. For both $Ru_1Ru_2$ and $Ru_2Ru_3$ sites empty in the outgoing cell, we calculate O and CO diffusion barriers of 1.25 eV and 1.87 eV, respectively. Barriers for other occupations of these two sites are then derived by correcting these values according to the changes in the adsorbate binding energy summarized in Table 1.

Table 3. LH CO oxidation reaction barriers (in eV). The different columns indicate the occupation of the third site within the surface unit-cell (empty (e), O or CO). The variation over the different columns thus reflects the lateral interactions with nearby adsorbed species.

| Reaction 3$^{rd}$-site | e | O | CO |
|---|---|---|---|
| $O@Ru_1Ru_2 + CO@Ru_2$ | 0.97 | 1.19 | 0.98 |
| $CO@Ru_1Ru_2 + O@Ru_2$ | 1.02 | 1.01 | 0.71 |
| $O@Ru_2 + CO@Ru_2Ru_3$ | 1.13 | 1.01 | 0.93 |
| $CO@Ru_2 + O@Ru_2Ru_3$ | 1.70 | 1.20 | 1.11 |

Depending on the occupation of the third site within the surface unit-cell, there are twelve possibilities for LH-type CO oxidation reactions with CO and O sitting in nearest-neighbor sites. Table 3 compiles the corresponding reaction barriers, again explicitly calculated for every configuration. Also in this case, significant lateral interactions can be discerned, although at a somewhat reduced level compared to the reactant binding energetics. ER-type reactions of an impinging CO molecule can in principle occur with a surface O atom adsorbed in any of the three sites. However, when the (on average for CO most attractive, cf. Table 1) $Ru_2$ site is empty, our PES scans showed that adsorption into this site is more favorable compared to an ER-reaction with O atoms either at $Ru_1Ru_2$ or $Ru_2Ru_3$ sites. We therefore only consider ER-processes either with O at the $Ru_2$ site (calculated barrier: 0.54 eV), or with O at the other two sites whenever the $Ru_2$ site is occupied by O or CO. The barriers for the latter cases are then, 0.22 eV (O@$Ru_1Ru_2$ with CO@$Ru_2$), 0.49 eV (O@$Ru_2Ru_3$ with CO@$Ru_2$) and 0.42 eV (O@$Ru_2Ru_3$ with O@$Ru_2$). To fix the prefactors for the ER rate constants estimates for the sticking coefficients are required.[12] In contrast to the non-activated $O_2$ adsorption, these sticking coefficients have to account for the significant reduction of CO entropy when passing through the tight transition state. We specifically choose a sticking coefficient of 0.05%, which roughly corresponds to a loss of 90% of the CO gas-phase entropy at the transition state. This particular choice leads to a prefactor that is one order of magnitude higher than the one employed by Hirvi et al.[26], who assumed a complete loss of entropy at the transition state. As further discussed below, even with our larger prefactor the ER reaction processes do not play a significant role around ambient pressure conditions and 600 K. In fact, for this to happen, the prefactor would need to be increased by another 1-2 orders of magnitude. For the present purposes the uncertainty in the ER prefactor is therefore not problematic. Future work will, however, aim for a more precise determination of prefactors for ER reactions in general, as we find the contribution of ER reactions increased at lower temperatures.

*2.3 1p-kMC simulation setup*
All 1p-kMC simulations are carried out with the kmos framework[27]. We employ a simulation cell comprising (20x20) surface unit-cells and periodic boundary conditions. Test simulations involving larger lattices produced identical average steady-state coverages and TOFs. Simulations are run for fixed ($T$,$p_{O2}$,$p_{CO}$)-conditions. After an initial equilibration period, steady-state values are obtained as long-time averages over 5 x $10^8$ 1p-kMC steps. The obtained steady states were always found to be independent from the initial starting configuration.

**3 RESULTS and DISCUSSION**

*3.1 CO and O adsorption geometries and energetics*
At the surface, Ru atoms can exhibit new electronic configurations due to a reduction of their sixfold O coordination in rutile bulk. Specifically, the O-poor $RuO_2$(111) surface termination used as basis for our 1p-kMC approach exhibits two types of fivefold coordinated surface Ru atoms ($Ru_1$ and $Ru_3$) and one threefold coordinated surface Ru atom ($Ru_2$) per surface unit-cell, cf. Fig. 1. In comparison, the analogue O-poor termination used in previous such work for the $RuO_2$(110) surface[11-14] has two symmetry-equivalent fourfold coordinated surface Ru atoms ($Ru_{br}$) and one fivefold coordinated surface Ru atom ($Ru_{cus}$) per surface unit-cell. Identifying similarities and differences for O and CO adsorption at these under-coordinated sites forms a general basis for an analysis of a possible structure sensitivity[17,18] of the CO oxidation reaction. At oxide surfaces it is furthermore instructive in view of relations to homogeneous catalysis at metal complexes.[28]

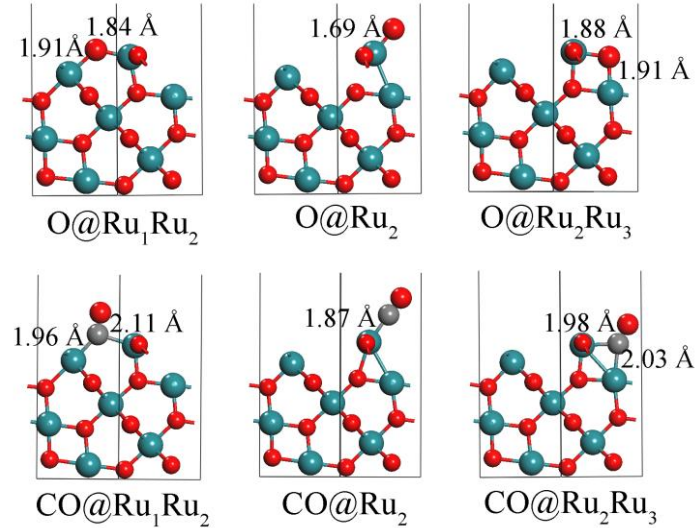

Figure 2. Side views of the adsorption geometries of O (upper panels) and CO (lower panels) at the three adsorption sites ($Ru_1Ru_2$, $Ru_2$, $Ru_2Ru_3$) offered by the O-poor $RuO_2(111)$ termination used as basis for the 1p-kMC approach. Ru atoms are shown as large blue spheres, O atoms as small red spheres, and C atoms as small gray spheres.

Figure 2 summarizes the calculated adsorption geometries of O and CO at the three different $RuO_2(111)$ binding sites. This immediately reveals that a classification of adsorption properties merely based on the level of under-coordination of the involved metal center falls short at the extended surfaces. At the $RuO_2(110)$ surface, both O and CO adsorption at the fivefold coordinated $Ru_{cus}$ atoms occurs in an atop position, with only one predominant adsorbate-substrate bond formed[5,7,29,30]. In contrast, at $RuO_2(111)$ both fivefold coordinated $Ru_1$ and $Ru_3$ atoms yield to bridge-type adsorption geometries that also involve the adjacent threefold coordinated $Ru_2$ atom. The O and CO bonding at the resulting $Ru_1Ru_2$ and $Ru_2Ru_3$ sites may therefore rather bear similarities to the equally bridge-type O and CO bonding at the fourfold coordinated $Ru_{br}$ atoms of the $RuO_2(110)$ surface. The Ru-O bond lengths shown in Fig. 2 are indeed more comparable in this respect. They are 1.69 Å at the $Ru_2$ site and in the range 1.84 – 1.91 Å at both $Ru_1Ru_2$ and $Ru_2Ru_3$ sites. This contrasts with 1.70 Å for atop $Ru_{cus}$-O and 1.91 Å for bridge $Ru_{br}$-O.[29] In the case of CO this translates to bond lengths of 1.87 Å at the $Ru_2$ site and 1.96 – 2.11 Å at $Ru_1Ru_2$ and $Ru_2Ru_3$ vs. 1.95 Å for atop $Ru_{cus}$-CO and 1.99 – 2.06 Å for bridge $Ru_{br}$-CO[30]. The strong asymmetry in the bridge site adsorption at $RuO_2(111)$ is thereby likely caused by the coordination to one fivefold ($Ru_1$,$Ru_3$) and one threefold ($Ru_2$) coordinated Ru atom. However, an asymmetric adsorption geometry has also been reported for high-coverage $CO_{br}$ adsorption in the bridge site coordinating to the two symmetry-equivalent $Ru_{br}$ atoms.[31] The geometric classification in terms of site-type rather than degree of under-coordination of the involved Ru atom(s) also carries over to the atop-type adsorption of O and CO at the $Ru_2$ site. Here, the bond lengths shown in Fig. 2 compare very well to the equivalent ones for atop adsorption at $Ru_{cus}$ atoms, despite the differences in the $Ru_{cus}$ (fivefold) and $Ru_2$ (threefold) coordination.

Table 4. O binding energy (in eV) at the three RuO$_2$(111) adsorption sites. The different columns indicate the occupation of the other two sites within the surface unit-cell (empty (e), O or CO). The variation over the different columns thus reflects the lateral interactions with nearby adsorbed species.

| | O binding energy | | | | | | | | |
|---|---|---|---|---|---|---|---|---|---|
| | e e | e O | e CO | O e | O O | O CO | CO e | CO O | CO CO |
| O@Ru$_1$Ru$_2$ | -2.32 | -2.41 | -2.83 | -1.24 | -1.45 | -1.21 | -1.45 | -1.83 | -2.09 |
| O@Ru$_2$ | -1.71 | -1.84 | -1.72 | -0.63 | -0.88 | -0.10 | -0.64 | -1.18 | -0.02 |
| O@Ru$_2$Ru$_3$ | -1.37 | -1.50 | -1.39 | -1.46 | -1.71 | -1.77 | -1.25 | -1.79 | -1.31 |

Proceeding to the adsorption energetics Tables 1 and 4 compile the calculated CO and O binding energies at the three adsorption sites, respectively. These energies exhibit a large variation of in parts up to almost 2 eV depending on the occupation of the other two sites in the surface unit-cell. The thereby implied strong lateral interactions prohibit any clear-cut qualitative distinction of the three adsorption sites. This is in strong contrast to the RuO$_2$(110) surface, where only small to negligible lateral interactions below ~0.2 eV were found between adsorbates at the two prominent adsorption sites.[11,12,30] We attribute this difference primarily to the high structural flexibility of the highly under-coordinated Ru$_2$ atom in comparison to the more rigid arrangement of the under-coordinated Ru atoms in the less open RuO$_2$(110) surface. Depending on the occupation of the nearby adsorption sites we calculate maximum displacements of the Ru$_2$ atom of up to 0.80 Å away from its relaxed position at the clean surface termination. These large relaxations are also apparent in the calculated adsorption geometries shown in Fig. 2 and contrast maximum relaxations calculated for the Ru$_{br}$ and Ru$_{cus}$ atoms at RuO$_2$(110) of the order of 0.1-0.2 Å.[29-31]

In case of the RuO$_2$(110) surface the small lateral interactions allowed to unambiguously distinguish in particular between O adsorption at Ru$_{br}$ (binding energy: ~2.4 eV [12,30]) and at Ru$_{cus}$ (binding energy: ~1 eV [12,30]). With a too strong O adsorption at Ru$_{br}$, this then immediately pointed at a prominent role of the cus site for steady-state CO oxidation at near-ambient and near-stoichiometric feed conditions.[5,7] Such a fingerprinting is not possible for the three RuO$_2$(111) sites. It would not even be possible on the basis of the actually calculated LH CO oxidation reaction barriers, cf. Table 3. Intriguingly, these barriers exhibit significantly smaller variations with the occupation of the third adsorption site than the concomitant reactant binding energies. This shows that approximate treatments of lateral interactions applied successfully at other surfaces[23-25] would not work at RuO$_2$(111) and an explicit full calculation as done here is required. Nevertheless, despite these smaller variations in the reaction barriers it is still not *a priori* obvious which reaction mechanism could possibly dominate the catalytic activity. Even more as at surfaces without such strong lateral interactions, this dictates to explicitly evaluate the interplay of the elementary processes within a microkinetic model in order to capture and analyze the catalytic function of this surface.

With respect to structure sensitivity, this data already indicates that the spatial arrangement of the active sites is another crucial factor for oxide catalysis that leads to a structure sensitivity of a reaction. Intriguingly, even in a hypothetical "low-coverage" limit, i.e. in the absence of lateral interactions, the calculated binding energies and reaction barriers reveal a structure sensitivity of the CO oxidation reaction at RuO$_2$(111). Despite the afore discussed similarity in the bonding geometries at the bridge-type and atop-type sites, the O and CO binding energies at like sites at the RuO$_2$(110) and RuO$_2$(111) surfaces do not compare, cf. columns ee without neighboring adsorbates in Tables 1 and 4 for RuO$_2$(111) with the above quoted binding energies at RuO$_2$(110) br and cus sites. Since the same adsorption site types (br, atop) at RuO$_2$(110) and RuO$_2$(111) involve differently coordinated surface Ru atoms, one could try to attribute this difference to an electronic effect arising from the differing degree of undercoordination[17]. However, even the O and CO binding energies at the Ru$_1$Ru$_2$ and Ru$_2$Ru$_3$ sites themselves differ by ~1 eV and ~0.6 eV, respectively. Both sites have an equivalent geometry (bridge) and electronic configuration (coordination to one threefold and one fivefold coordinated Ru atom). This demonstrates that the concepts that have been put forward to classify structure sensitivity at metal catalysts[17] cannot be carried over to these oxide surfaces.

*3.2 Coverage and turnover frequency at catalytically active conditions*

We concentrate our analysis of the catalytic function on a reaction temperature of 600 K, which lies at the upper range of interest for CO oxidation. This temperature was also specifically studied in the previous 1p-kMC work at $RuO_2(110)$ [11-14], which then enables the detailed comparison targeted here.

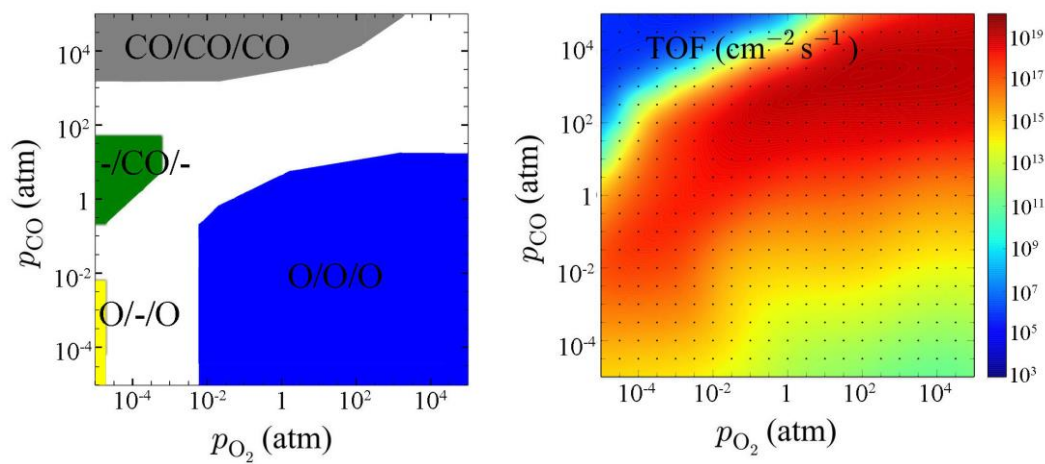

Figure 3. 1p-kMC computed steady-state O and CO surface coverage (left panel) and CO turnover frequency, TOF (right panel). Shown are data for a range of CO and oxygen partial pressures at 600 K. The short-hand notation for the various differently colored coverage phases in the left panel indicates the dominant species (O, CO, or empty "–") at the three sites offered by the $RuO_2(111)$ surface, e.g. O/O/O/ indicates a steady-state O coverage >80% at sites $Ru_1Ru_2/Ru_2/Ru_2Ru_3$. White regions between the colored phases indicate coexistence of species at least at one of the three sites.

Figure 3 displays the 1p-kMC calculated steady-state average surface coverages and CO oxidation TOFs over a range of reaction partial pressures around ambient conditions. At the lowest $p_{CO}$~$10^{-5}$ atm shown the catalytic activity starts to die out and the surface coverages obtained as a function of oxygen pressure necessarily agree with those obtained within the constrained *ab initio* thermodynamics approach[29,30], which neglects any kinetic effects of ongoing catalytic reactions on the surface composition. The correspondingly calculated surface phase diagram is shown in Fig. 4 and equally exhibits a transition between a phase where the $Ru_1Ru_2$ and $Ru_2Ru_3$ sites are covered with oxygen atoms (denoted as O/–/O) and a fully O-covered surface (denoted as O/O/O). Within the constrained *ab initio* thermodynamics approach configurational entropy is neglected. The phase transition is therefore infinitely sharp, whereas the 1p-kMC simulations fully account for the temperature-induced widening of the transition over a finite range of oxygen pressures.[11,12,15]

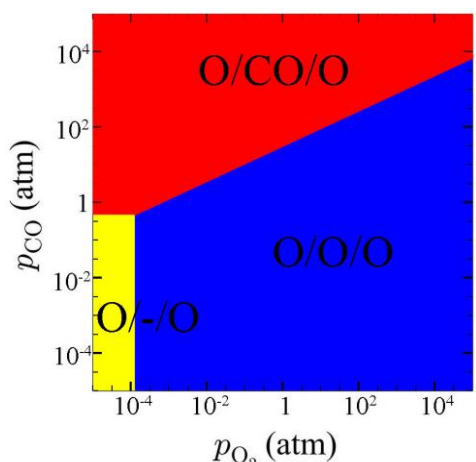

Figure 4. Steady-state coverage map as obtained from constrained *ab initio* thermodynamics. The nomenclature for the different phases is the same as in the left panel of Fig. 3.

At increasing CO pressures catalytic activity sets in. The higher CO impingement increases the probability for ER reactions and enables LH reactions due to the increased stabilization of CO at the surface. As shown in Fig. 4, from a thermodynamic point of view this stabilization should notably set in close to ambient CO pressures at low $p_{O_2}$. With increasing oxygen pressures correspondingly higher $p_{CO}$ are then required to lead to the corresponding O/CO/O phase, in which CO predominantly covers the $Ru_2$ sites. Comparing these predictions to the actual 1p-kMC results in Fig. 3 we indeed start to find a significant CO concentration at the surface in this pressure range. Also, the thermodynamically intuitive shift of the corresponding coexistence range (depicted as a white region in Fig. 3) to higher $p_{CO}$ with increasing $p_{O_2}$ is obtained. However, strong kinetic effects lead to a completely different surface composition as anticipated by the approximate constrained *ab initio* thermodynamics theory. While the dominant CO coverage at the $Ru_2$ site is retained, these kinetic effects strongly suppress the presence of oxygen at the $Ru_1Ru_2$ and $Ru_2Ru_3$ sites. Instead of the thermodynamically predicted O/CO/O phase, these two sites are thus either largely empty (the region denoted –/CO/– in Fig. 3) or at higher $p_{CO}$ largely covered with CO (the CO/CO/CO region in Fig. 3). The consideration of the kinetic effects on the surface population due to the ongoing CO oxidation reactions therewith leads to a much earlier CO poisoning of the surface with increasing CO pressures. This finding was analogously obtained in the earlier work on the $RuO_2(110)$ surface; compare specifically with the $CO_{br}$/– and $CO_{br}/CO_{cus}$ regions in the upper left part of Fig. 6 in Ref. [12], which is completely equivalent to the present Fig. 4 for $RuO_2(111)$. Here and there, the central reason for this strong suppression of surface O species at corresponding high CO partial pressures are the kinetic limitations in finding two adjacent empty sites required for the dissociative $O_2$ adsorption at the highly covered surfaces.[14]

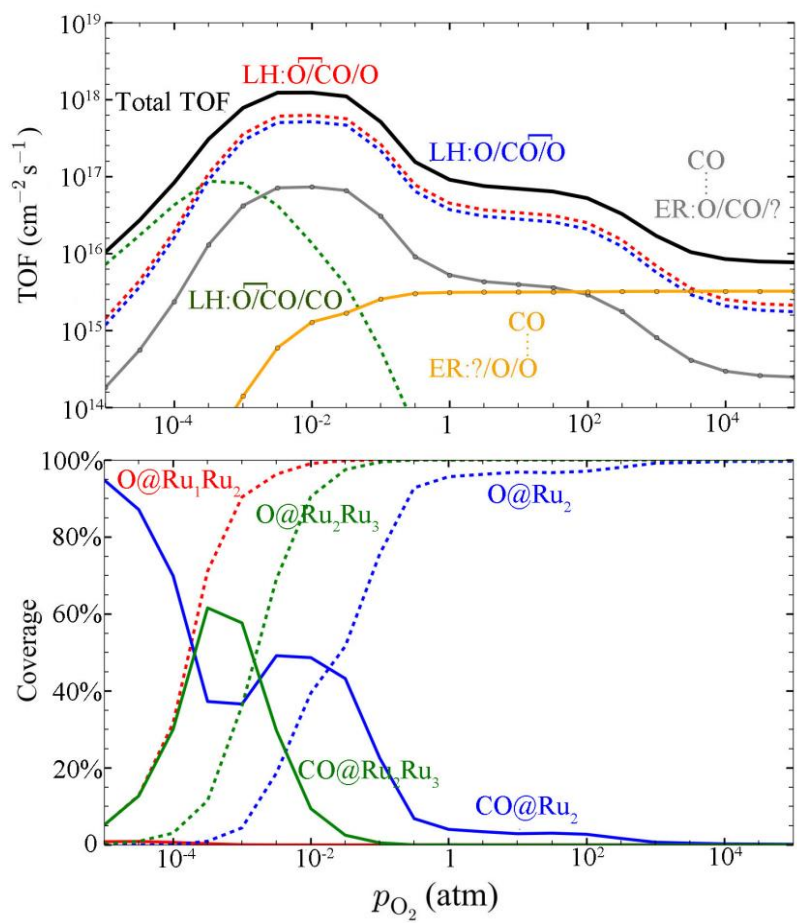

Figure 5. 1p-kMC computed steady-state turnover frequency, TOF (upper panel), and surface coverages (lower panel) at $p_{CO}$ = 1 atm and 600 K. In addition to the total TOF, the individual contributions from different reaction mechanisms are shown as colored lines. The employed short-hand notation indicates the predominant coverage at the three sites as in Figs. 3 and 4, with a bracket above two sites indicating the two reaction intermediates involved in LH reactions, and a CO atop one of the sites indicating which O adsorbate is picked up in ER reactions.

Corresponding pressure regions in the upper left part of Fig. 3 which presently yield CO-rich surface compositions are in reality prone to lead to further oxide reduction. This is not possible in the present 1p-kMC model which focuses on the surface adsorption sites at an otherwise intact oxide surface. In the following we therefore rather concentrate on relatively lower CO pressures, where at least for sufficiently large $p_{O2}$ Fig. 3 predicts an appreciable concentration of surface O species and where we would therefore at least expect a metastability of the underlying oxide matrix. Specifically, Fig. 5 shows the detailed surface coverages and TOF contributions of individual reaction mechanism for $p_{CO}$ = 1 atm and varying oxygen pressures. The peak activity obtained for these conditions is about one order of magnitude lower than the peak activity observed for RuO$_2$(110), which was 4 x 10$^{19}$ cm$^{-2}$s$^{-1}$.[12,13] It is 1 x 10$^{18}$ cm$^{-2}$s$^{-1}$ along the plotted line of $p_{CO}$ = 1atm in Fig. 5. The overall maximum activity for the RuO$_2$(111) facet at 600 K is 6 x 10$^{19}$ cm$^{-2}$s$^{-1}$ in Fig. 3, which is thus almost identical to the peak activity of RuO$_2$(110).

The two dominantly contributing mechanisms to this peak activity are the LH mechanisms O@Ru$_1$Ru$_2$ + CO@Ru$_2$ (with O@Ru$_2$Ru$_3$, barrier: 1.19eV) and O@Ru$_2$Ru$_3$ + CO@Ru$_2$ (with O@Ru$_1$Ru$_2$, barrier: 1.20eV). At the pressures corresponding to this peak activity, these two mechanisms benefit mostly from the prevalent surface composition, which is characterized by a coexistence of oxygen and CO at the $Ru_2$ and $Ru_2Ru_3$ sites, with O atoms already predominantly present at the $Ru_1Ru_2$ sites (white coexistence region in Fig. 3). ER reactions do not contribute significantly to the peak activity. They only start to take over at very high oxygen pressures, cf. Fig. 5. At these extremely oxygen-rich conditions the surface is completely oxygen poisoned (O/O/O phase). The concomitantly low surface CO concentration then effectively suppresses any of the LH mechanisms. This increased contribution of ER processes to the total TOF at high $p_{O2}$ leads to the rather weak decline of the catalytic activity towards the right in the TOF-map of Fig. 3. This contrasts the plummeting TOFs for increasing $p_{O2}$ obtained in the previous 1p-kMC work on $RuO_2$(110), where ER processes were not considered.[11-13] Whether the latter works need to be revised in this respect, or whether the current ER contribution is overestimated depends critically on the employed prefactor in the corresponding first-principles rate constants. Compared to the work of Hirvi et al.[26] the estimate used in this work is an upper bound and even then ER processes do not play a role for the near-ambient peak activity at 600 K.

Similar to the case for $RuO_2$(110)[12,30] it is thus by far not the LH mechanisms that exhibit the lowest reaction barriers that dominate the catalytic peak activity around ambient pressures. In principle the $RuO_2$(111) surface would exhibit mechanisms with barriers even about 0.5 eV lower, cf. Table 3. Yet, they cannot contribute in the interplay of all elementary processes, which underscores the importance of the explicit evaluation of this interplay within a microkinetic model. Interestingly, the shortcoming of individual energy barrier values to predict catalytic activity extends also to the comparison of the two $RuO_2$ facets. The two dominant reaction mechanisms at $RuO_2$(111) exhibit barriers that are about 0.3 eV higher than the barrier of the dominant reaction mechanism at the $RuO_2$(110) surface ($O_{cus}$ + $CO_{cus}$, 0.8 eV [12,30]). In a naïve Arrhenius picture and with $RuO_2$(110) and $RuO_2$(111) exhibiting approximately equal site densities per area one would then expect the two peak catalytic activities to differ by ~exp (-0.3 eV / $k_BT$) ≈ $10^{-3}$ at 600K. In contrast, the explicit 1p-kMC simulations yield virtually identical peak activities. In addition to the differing absolute TOF value, the computed peak activity of $RuO_2$(111) occurs furthermore at different partial pressure ratios than the one at $RuO_2$(110). For the $p_{CO}$ = 1atm condition in Fig. 5 it is obtained at $p_{O2}$ = 6.8 x $10^{-3}$ atm, i.e. for a partial pressure ratio of $p_{CO}/p_{O2}$ ~ 150. In contrast, for $RuO_2$(110) at corresponding near-ambient total pressures it was obtained for a partial pressure ratio of $p_{CO}/p_{O2}$ ~ 5.[12] The catalytic activity of the $RuO_2$(111) surface extends therefore much more to reducing feed conditions. This is fully consistent with recent experimental reports of a preferential reduction of the apical $RuO_2$(111) facets of $RuO_2$ crystals in reducing feeds.[9]

Finally, we return again to the exceeding similarity of the peak activities found for $RuO_2$(110) and $RuO_2$(111) at 600 K. First of all, this is already surprising in view of the differences in the elementary processes and their rate constants discussed in Section 3.1. It is even more surprising when considering the two quite different reasons why the catalytic activity at the two $RuO_2$ facets cannot be captured with prevalent mean-field kinetic models. At $RuO_2$(110) this arises out of a strong binding of oxygen at the br adsorption sites[12,30], which restricts the catalytic activity primarily to the remaining cus sites. Even though there are only insignificant lateral interactions between reaction intermediates at these sites, the row-like arrangement of the cus sites together with concomitant diffusion limitations in the resulting one-dimensional cus-trenches then lead to the non-random spatial distribution of the reaction intermediates that causes the break-down of the mean-field assumptions.[13,14] In contrast, at $RuO_2$(111) there is no extended site network, but instead independent tri-site clusters to which the essential elementary processes are confined. Among this group of sites it is then strong lateral interactions that leads to site occupation and activity patterns that are beyond the reach of mean-field kinetics.

In view of these qualitative differences in the site arrangement, interplay of elementary processes and even the underlying individual elementary processes, CO oxidation at these two $RuO_2$ facets appears as a structure sensitive reaction *par excellence*. In this respect, the almost identical peak activity albeit at differing partial pressure ratios has another important implication.

For partial pressure ratios $5 < p_{CO}/p_{O2} < 50$, i.e. in between the limits where one or the other facet exhibits its peak activity at near-ambient pressures, both facets correspondingly exhibit somewhat smaller, but still high activity. Precisely for such partial pressure ratios that are most relevant for practical catalysis and that are concomitantly typically explored in experimental studies, we can therefore easily find multiple absolute pressure conditions where both facets again exhibit identical TOFs. This should be seen with respect to the traditional classification or definition of a reaction as structure insensitive based merely on the macroscopically observed catalytic function as typically explored only over a small set of feed conditions.[1-3] As exemplified by the data obtained here for $RuO_2(111)$ and $RuO_2(110)$ this can be a dangerous concept that does not adequately capture the underlying micro- to mesoscopic complexities. We stress, however, that this is a general statement based on the well-defined theory-theory comparison of the two 1p-kMC models of $RuO_2(110)$ and $RuO_2(111)$. For Ru nanoparticles in oxidizing feeds Hoon Joo et al.[32] in fact reported a structure sensitivity for the CO oxidation reaction, which they, however, ascribed to a varying degree of oxidation with nanoparticle size.

## 4 CONCLUSIONS

We presented a 1p-kMC study of CO oxidation at $RuO_2(111)$ and compared the obtained detailed data on the adsorption energetics and geometry, as well as surface composition and catalytic activity under steady-state reaction conditions with corresponding data available for the $RuO_2(110)$ facet. This comparison provides detailed insights into the structure sensitivity of this reaction and on the catalytic function of $RuO_2$ nanoparticles. In terms of under-coordinated surface Ru atoms both facets share similarities at first sight. A more detailed inspection of the adsorption sites reveals, however, that the bonding of the reaction intermediates at the fivefold coordinated Ru atoms present at both surfaces is qualitatively different. Some structural similarity can instead be discerned on the level of similarly coordinated adsorption sites, in particular bridge-type sites present at both facets. Notwithstanding, even at these sites the adsorbate binding energetics is largely different, which shows that the understanding of structure sensitivity in terms of geometric and electronic factors that has been developed for metal catalysts does not carry over to these oxide surfaces. This even more so, as at $RuO_2(111)$ the presence of a structurally most flexible, only threefold coordinated surface Ru atom leads to strong lateral interactions. These interactions in fact prevent any straightforward identification of particularly "active" sites – a concept that has been so successfully applied for the "cus"-sites of the extensively studied $RuO_2(110)$ surface.

Also on the level of the spatial arrangement of the individual adsorption sites both facets exhibit qualitative differences. At $RuO_2(111)$ the three adsorption sites situated within one surface unit-cell are largely decoupled from sites in neighboring cells. This together with the strong lateral interactions between adsorbates at the three clustered sites yields a molecular-type catalytic behavior that cannot be grasped with prevalent mean-field microkinetic models. In contrast, at $RuO_2(110)$ only very modest lateral interactions seemingly suggest the applicability of mean-field approaches. Here, however, the strong O binding at one of the two adsorption site types largely poisons the corresponding sites at near-stoichiometric feeds, which then leads to a micropatterning of the surface. In consequence, the catalytic reactions run prominently along one-dimensional trenches, which – again – is beyond the reach of mean-field kinetics.

Despite all of these differences, at 600 K the peak activity of both facets is virtually identical. The underlying surface coverages and concomitant reaction patterns are thereby quite different though. With the surface composition varying differently with reactant pressures, the peak activity of the two facets is correspondingly obtained at different partial pressure ratios. Already the data obtained for these two $RuO_2$ facets thus suggests that the concept of one set of "optimum reaction conditions" is generally short-sighted for catalyst nanoparticles. For near-ambient pressures and near-stoichiometric feeds both facets exhibit slightly lower activities compared to their respective peak activities. Intriguingly, these activities are, however, again exceedingly similar to each other. Probing the catalytic activity only for a restricted set of gas-phase conditions in this range would therefore erroneously suggest the reaction to be structure insensitive – at least according to the prevailing macroscopic definition of structure sensitivity.

At the elevated temperature analyzed in this study, the peak activity of $RuO_2(111)$ is shifted to lower $O_2$ pressures compared to $RuO_2(110)$. This seems consistent with recent experimental reports pointing at a prominent role of these apical facets in the reduction of $RuO_2$ crystals at elevated temperatures. As to the long-term steady-state activity, the obtained different composition and activity patterns of the two studied facets hint at interesting mass transport effects over the facet edges of $RuO_2$ nanoparticles. These will be the focus of ensuing work along our long-term track of systematically bridging between single-crystal model and real catalysis.

**AUTHOR INFORMATION**

Corresponding Author
*E-mail: karsten.reuter@ch.tum.de.
Notes
The authors declare no competing financial interest.

**ACKNOWLEDGEMENTS**

We gratefully acknowledge generous access to computing time at the Leibniz Supercomputing Centre under grant 'pr86de'. T.W. is thankful for support from the China Scholarship Council and acknowledges enlightening discussions with Dr. Farnaz Sotoodeh, Dr. Mie Andersen and Juan Manuel Lorenzi.